\documentclass[a4paper]{jpconf}
\usepackage{graphicx}

       \newcommand{\Cc}{ {\mathcal{C}} }

       \newcommand{\Hc}{ {\mathcal{H}} }
       
       \newcommand{\Jc}{ {\mathcal{J}} }

       \newcommand{\Pc}{ {\mathcal{P}} }

       \newcommand{\Xc}{ {\mathcal{X}} }
       
       \newcommand{\Zc}{ {\mathcal{Z}} }
       \newcommand{\bv}{{\mathbf v}}

       \newcommand{\bA}{{\mathbf A}}
       \newcommand{\bB}{{\mathbf B}}

       \newcommand{\bJ}{{\mathbf J}}

  \newcommand{\nn}{\nonumber}
  \newcommand{\bOm}{{\mathbf \Omega}}
  
  \newcommand{\nb}{\nabla}
  
  \newcommand{\wrm}{\mathrm{w}}

\usepackage{amssymb}
 \usepackage{amsthm}
 \expandafter\let\csname equation*\endcsname\relax

\expandafter\let\csname endequation*\endcsname\relax
 \usepackage{amsmath}

 \usepackage{mathtools}
\usepackage{amsfonts}
\usepackage{graphicx}% Include figure files
\usepackage{dcolumn}% Align table columns on decimal point
\usepackage{bm}% bold math
\usepackage[export]{adjustbox}
\usepackage{wrapfig}
\usepackage{lipsum}
\usepackage[titletoc,toc,title]{appendix}
\usepackage[colorlinks]{hyperref}

\begin{document}
\title{A discrete Nambu bracket for 2D extended Magnetohydrodynamics}

\author{D A Kaltsas$^1$, M Kraus$^{2,3}$ and G N Throumoulopoulos$^1$}

\address{$^1$ Department of Physics, University of Ioannina, GR 451 10 Ioannina, Greece}
\address{$^2$ Max-Planck-Institute for Plasma Physics, Garching, Germany }
\address{$^3$ Technische Universit\"at M\"unchen, Zentrum Mathematik, Garching, Germany}

\ead{dkaltsas@cc.uoi.gr}

\begin{abstract}
In this note we propose a trilinear bracket formulation for the Hamiltonian extended Magnetohydrodynamics (XMHD) model with homogeneous mass density. The corresponding two-dimensional representation is derived by performing spatial reduction on the three-dimensional bracket, upon introducing a symmetric representation for the field variables. Subsequently, the trilinear bracket of the resulting two-dimensional, four-field model is discretized using a finite difference scheme, which results in semi-discrete dynamics that involve the Arakawa Jacobian. Simulations of planar dynamics show that this scheme respects the desired conservation properties to high precision.  
\end{abstract}
  \vspace{-2mm}
\section{Introduction}
The Hamiltonian formulation of ideal fluid models \cite{Morrison1998} (e.g. Magnetohydrodynamics) in Eulerian viewpoint involves noncanonical variables and Poisson operators that are degenerate and inhomogeneous in phase space $\Pc$. The degeneracy of the Poisson operator results in the emergence of topological invariants, the so-called Casimirs. Unlike the Hamiltonian functional, being conserved as a  consequence of the necessary property of antisymmetry of the associated Poisson bracket ($d\Hc/dt=\{\Hc,\Hc\}=0$), the Casimirs are conserved due to its degeneracy, i.e.
\begin{eqnarray}
\{\Cc,F\}=0\,,\quad \forall F\in C^\infty(\Pc)\,. \label{casimir_determining}
\end{eqnarray}
This means that, the Casimir $\Cc$ permutes with any functional $F$, not only with the Hamiltonian $\Hc$ of the system. In this regard, one could state that the conservation of $\Cc$'s is more fundamental.

 In \cite{Nevir1993} and later on in \cite{Blender2015}, an infinite dimensional generalization of the  so-called Nambu bracket \cite{Nambu1973} was introduced to describe inviscid three and two dimensional hydrodynamics. This new formulation brings out the hidden conservation of the Casimir on the same level with the Hamiltonian through a trilinear form $[C^\infty(\Pc)]^{\otimes 3}\rightarrow C^\infty(\Pc)$, which is completely antisymmetric in its three arguments and incorporates a second Hamiltonian functional for the description of the dynamics. The second Hamiltonian is actually the Casimir invariant, i.e. the fluid helicity and enstrophy for 3D and 2D flows, respectively. Therefore, in view of this structure, both the Hamiltonian functional and the Casimir invariant are conserved due to antisymmetry. 
%Bringing out the hidden conservation of the Casimir on the same level with the Hamiltonian, reasonably would imply certain modifications regarding the geometric properties of the field-theoretic Nambu structure in comparison with the corresponding Poisson structure, which have not been elucidated on a satisfactory level until now. However, despite the obscure status quo in this field, t
This property has been successfully exploited for the construction of energy-Casimir preserving algorithms in the context of fluid dynamics. As Salmon suggested in \cite{Salmon2005,Salmon2007}, upon discretizing the trilinear bracket in such a way that the antisymmetry property is preserved, the conservation of the discrete counterparts of the two Hamiltonians is ensured by simple algebraic cancellations. Therefore, the resulting discretized equations will retain the desired conservation properties, which is expected to be beneficial in decreasing the error of the numerical solution and enhancing  the stability of the simulations. 

With this note we aspire to extend this concept in the case of extended Magnetohydrodynamics (XMHD), that is a quasineutral two-fluid model that incorporates electron inertial effects and has a Hamiltonian structure in its ideal limit \cite{Abdelhamid2015,Lingam2015}. Our main motivation for studying XMHD is that the inclusion of Hall drift and electron inertial effects results in a more realistic description for the plasmas than MHD provides and in addition it gives rise to fast reconnection, which is observed in extraterrestrial environments where the plasma is nearly collisionless and also in laboratory experiments, e.g. reconnection during the sawtooth oscillations occuring in Tokamaks. Suppressing artificial dissipation is crucial for ensuring the fidelity of reconnection simulations.
 \vspace{-2mm}
\section{Incompressible XMHD dynamics}
The equations of motion of incompressible XMHD dynamics in the vorticity representation are
\begin{eqnarray}
\partial_t\mathbf{\Omega}\hspace{-2mm}&=&\hspace{-2mm}\nabla\times\left(\bv\times\mathbf{\Omega}\right)+\nabla\times\left(\bJ\times\bB^*\right) \label{mom_omega}\,,\\
\partial_t\bB^*\hspace{-2mm}&=&\hspace{-2mm}\nabla\times\left(\bv\times\bB^*\right)-d_i\nabla\times\left(\bJ\times \bB^*\right)+d_e^2\nabla\times\left(\bJ\times\mathbf{\Omega}\right)\,,\label{induction}
\end{eqnarray}
where $d_i$ and $d_e$ are the normalized ion and electron skin depths, respectively,  $\bJ=\nabla\times \bB$, $\mathbf{\Omega}=\nabla\times \bv$ and
\begin{eqnarray}
\bB^*=\bB-d_e^2\Delta\bB \,.\label{bstar}
\end{eqnarray}
with $\Delta:=\nb^2$. The Hamiltonian structure of the barotropic version of this model has been identified in \cite{Abdelhamid2015}. Ignoring the compressible part of this formulation and also expressing it in vorticity representation we find 
that the Hamiltonian is given by 
\begin{eqnarray}
\Hc=\frac{1}{2}\int d^3x \left(\boldsymbol{\xi}\cdot\bOm+\bB^*\cdot \bB\right)\,,
\end{eqnarray}
where $\boldsymbol{\xi}$ is a vector potential, $\bv=\nb\times\boldsymbol{\xi}$, and the corresponding noncanonical Poisson bracket
\begin{eqnarray}
&&\{F,G\}=\int d^3x\,\big\{ (\nabla\times\bv)\cdot\left[(\nabla\times F_{\bOm} )\times(\nabla\times G_{\bOm})\right]\nn\\
&&\bB^*\cdot\left[(\nabla\times F_{\bOm})\times(\nabla\times G_{\bB^*})-(\nabla\times G_{\bOm})\times(\nabla\times F_{\bB^*})\right]\nn \\
&&-d_i\bB^*\cdot\left[(\nabla\times F_{\bB^*})\times(\nabla\times G_{\bB^*}) \right]+d_e^2(\nabla\times\bv)\cdot[(\nabla\times F_{\bB^*})\times(\nabla\times G_{\bB^*})]\big\}\,, \label{poisson_vort_repr}
\end{eqnarray}
with $F_u$ denoting the functional derivative of $F$ with respect to u. One can find two Casimirs satisfying \eqref{casimir_determining}, which are generalized helicities of the form
\begin{eqnarray}
\Xc=\int d^3x\, \bB^*\cdot\left(\bv-\frac{d_i}{2d_e^2}\bA^*\right)\,,\label{casimir1}\quad
\Zc=\frac{1}{2}\int d^3x\, \left[\bA^*\cdot\bB^*+d_e^2\bv\cdot\bOm\right]\,. \label{casimir2}
\end{eqnarray}
Here, $\bA^*$ is the generalized magnetic potential. In view of \eqref{casimir2}, it can be verified that the incompressible XMHD dynamics are described by 
\begin{eqnarray}
\partial_tF=\{F,\Hc,\Zc\}\,, \label{Hamilton_equation}
\end{eqnarray}
where
\begin{eqnarray}
&&\{F,G,K\}=\int_Vd^3x\,\bigg\{\frac{1}{3d_e^2}(\nabla\times F_\bOm)\cdot[(\nabla\times G_\bOm) \times (\nabla \times K_\bOm)]\nn \\
&&\hspace{-1cm}-(\nabla\times F_{\bB^*})\cdot\left[\frac{d_i}{3}(\nabla\times G_{\bB^*})\times (\nabla\times K_{\bB^*})
-(\nabla\times G_\bOm) \times(\nabla\times K_{\bB^*})\right]\bigg\}
+\circlearrowright (F,G,K)\,, \label{nambu}
\end{eqnarray}
is our trilinear bracket with $\circlearrowright(F,G,K)$ denoting cyclic permutation.
 \vspace{-2mm}
\section{Reduction to 2D}
A two-dimensional counterpart of this model can still capture the Hall and electron effects while being computationally more tractable than the 3D version. For this reason we employ a standard reduction to translationally symmetric dynamics upon decomposing the vector fields as follows
\begin{eqnarray}
\bOm=\omega(x,y,t)\hat{z}+\nabla \wrm(x,y,t)\times\hat{z}\,,\quad
\bB=b(x,y,t) \hat{z}+\nabla\psi(x,y,t)\times\hat{z}\,,\label{ts}
\end{eqnarray}
hence, from \eqref{bstar} one can be find $b^*=b-d_e^2\Delta b$ and $\psi^*=\psi-d_e^2\Delta\psi$. 
In view of the translationally symmetric decomposition \eqref{ts} the functional derivatives with respect to the vector fields $\bOm$ and $\bB^*$ can be expressed in terms of the functional derivatives with respect to the scalar fields $(\omega,\wrm,b^*,\psi^*)$ leading to
\begin{eqnarray}
\nabla\times F_{\bOm}=F_{\wrm} \hat{z}+\nabla F_{\omega}\times\hat{z} \,,\quad \nabla\times F_{\bB^*}=F_{\psi^*}\hat{z}+\nabla F_{b^*}\times\hat{z}\,, \label{curl_red_fun_der}
\end{eqnarray}
and consequently the trilinear bracket \eqref{nambu} reduces to 
 \begin{eqnarray}
 \{F,G,K\}=\int d^2x\,\Big\{\frac{1}{d_e^2}F_{\wrm}[G_{\omega},K_{\omega}]-d_iF_{\psi^*}[G_{b^*},K_{b^*}]\nn \\
 +F_{\psi^*}[G_{\omega},K_{b^*}]+F_{b^*}[G_{\omega},K_{\psi^*}]+F_{b^*}[G_{\wrm},K_{b^*}]\Big\}+\circlearrowright(F,G,K)\,, \label{reduced_Nambu}
 \end{eqnarray}
where $[f,g]:=(\partial_xf)(\partial_yg)-(\partial_xg)(\partial_yf)$. The translationally symmetric dynamics is governed by the four-field model
  \begin{eqnarray}
 &&\partial_t \wrm=[\chi,\wrm]+[b,\psi^*] \,, \label{ts_xmhd_dyn_1}\\
 &&\partial_t\psi^*= [\chi, \psi^*]-d_i[b,\psi^*]+d_e^2[b,\wrm]\,, \\ 
 &&\partial_t\omega=[\chi,\omega]+[\psi,\Delta \psi]-d_e^2[b,\Delta b]\\
 &&\partial_t b^* =[\chi,b^*]+[\wrm,\psi]-d_i[\psi,\Delta\psi]+d_id_e^2[b,\Delta b]+d_e^2[b,\omega]\,,\label{ts_xmhd_dyn_4}
 \end{eqnarray}
 which follow from \eqref{Hamilton_equation} and \eqref{reduced_Nambu} and the translationally symmetric $\Hc$ and $\Zc$.
 \section{Finite difference discretization}
  \begin{figure}
 \begin{center}
 \includegraphics[scale=0.25]{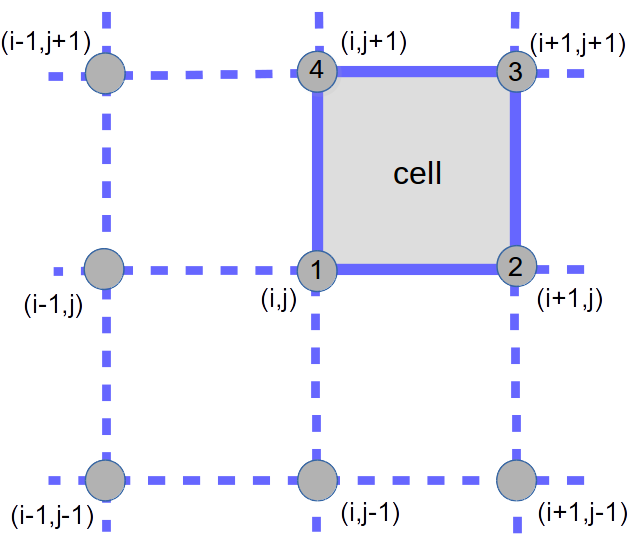}
  \end{center}
    \vspace{-2mm}
  \caption{A regular square grid with spacing $h$ in both directions is employed for a simple finite difference discretization of the bracket \eqref{reduced_Nambu}. \label{fig_1}}
    \vspace{-2mm}
 \end{figure}
 Following Salmon's algorithm \cite{Salmon2005}, to retain the antisymmetry property on the discrete level we impose the following symmetrization
 \begin{eqnarray}
 \int d^2x\, f[g,h]\longrightarrow \frac{1}{3}\left(\int d^2x\, f[g,h]+\int d^2x\, h[f,g]+\int d^2x\, g[h,f]\right)\,,\label{symmetrization}
 \end{eqnarray}
 due to the identity $\int d^2x\, f[g,h]=\int d^2x\, h[f,g]=\int d^2x\, g[h,f]$ and appropriate boundary conditions, e.g. periodic. Then it is sufficient to discretize the inner Jacobi-Poisson bracket so as to be antisymmetric, e.g.
 \begin{eqnarray}
 \int d^2x\, \frac{\delta F}{\delta a}\left[\frac{\delta G}{\delta b},\frac{\delta K}{\delta c}\right]\longrightarrow \frac{1}{8h^2}\sum_{cells} \left(\frac{\partial F}{\partial a_1}+\frac{\partial F}{\partial a_2}+\frac{\partial F}{\partial a_3}+\frac{\partial F}{\partial a_4}\right)\times\nn\\
 \times\left[\left(\frac{\partial G}{\partial b_3}-\frac{\partial G}{\partial b_1}\right)\left(\frac{\partial K}{\partial c_4}-\frac{\partial K}{\partial c_2}\right)-\left(\frac{\partial K}{\partial c_3}-\frac{\partial K}{\partial c_1}\right)\left(\frac{\partial G}{\partial b_4}-\frac{\partial G}{\partial b_2}\right)\right]\,, \label{discrete_Lie-Poisson}
 \end{eqnarray}
 resulting in a discrete analogue of \eqref{reduced_Nambu} $\{F,G,K\}_{d}$ (see figure \ref{fig_1} and also \cite{Salmon2005,Salmon2007}). The semi-discrete dynamics is described by the following system of ODEs
 \begin{equation}
 \frac{d F_{ij}(t)}{dt}=\{F_{ij},\Hc_{d},\Zc_{d}\}_{d}\,, \quad i,j=1,...,N\,, \label{semi_discrete}
 \end{equation}
where $\Hc_d$, $\Zc_d$ are the discrete analogues of $\Hc$ and $\Zc$, respectively. In view of \eqref{semi_discrete}, \eqref{reduced_Nambu} and \eqref{symmetrization}--\eqref{discrete_Lie-Poisson} we find 
%  \begin{eqnarray}
% &&\frac{dv_{ij}}{dt}=\Jc_{ij}(\chi,v)+\Jc_{ij}(b,\psi^*) \,, \label{discrete_ts_xmhd_dyn_1}\\
% &&\frac{d\psi^{*}_{ij}}{dt}= \Jc_{ij}(\chi, \psi^*)-d_i\Jc_{ij}(b,\psi^*)+d_e^2\Jc_{ij}(b,v)\,, \\ 
% &&\frac{d\omega_{ij}}{dt}=\Jc_{ij}(\chi,\omega)+\Jc_{ij}(\psi,\Delta \psi)-d_e^2\Jc_{ij}(b,\Delta b)\\
% &&\frac{db^{*}_{ij}}{dt} =\Jc_{ij}(\chi,b^*)+\Jc_{ij}(v,\psi)-d_i\Jc_{ij}(\psi,\Delta\psi)+d_id_e^2\Jc_{ij}(b,\Delta b)+d_e^2\Jc_{ij}(b,\omega)\,,\label{discrete_ts_xmhd_dyn_4}
% \end{eqnarray} 
that the semi-discrete dynamical equations are provided by \eqref{ts_xmhd_dyn_1}--\eqref{ts_xmhd_dyn_4} with the field variables on the lhs being replaced by their values at $(i,j)$ node and the Jacobians on the rhs by the Arakawa Jacobian $\Jc_{ij}$ as it is given in \cite{Arakawa1966}. Note that upon setting $d_i=d_e=0$ and $\wrm=b=0$, the reduced MHD model is retrieved from \eqref{ts_xmhd_dyn_1}--\eqref{ts_xmhd_dyn_4}. In this case our semi-discrete system is equivalent with the semi-discrete equations found in \cite{Kraus2016} where a Variational Integrator approach \cite{Kraus2015} was employed. 
 \begin{figure}
\centering
\includegraphics[scale=0.34]{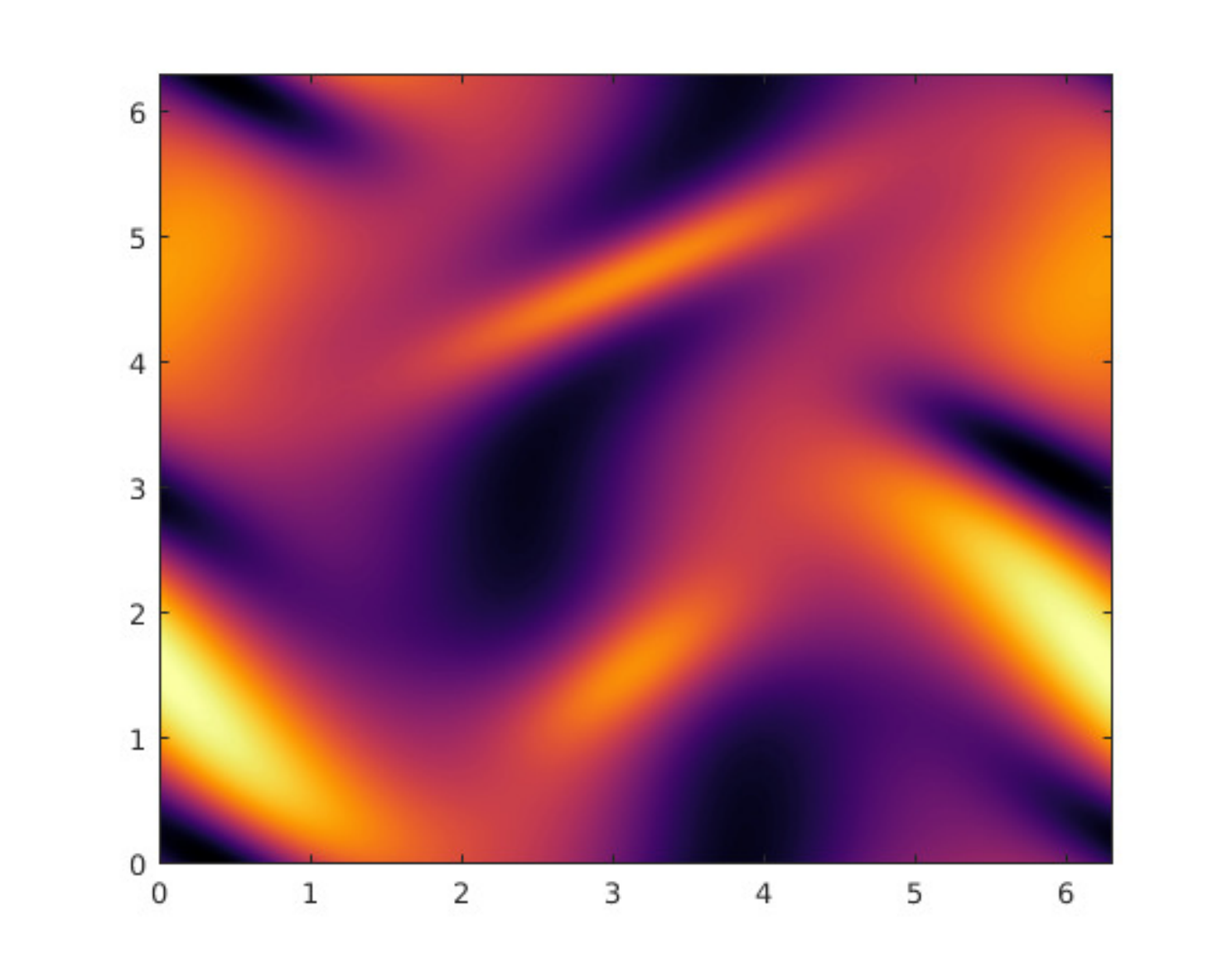}\includegraphics[scale=0.36]{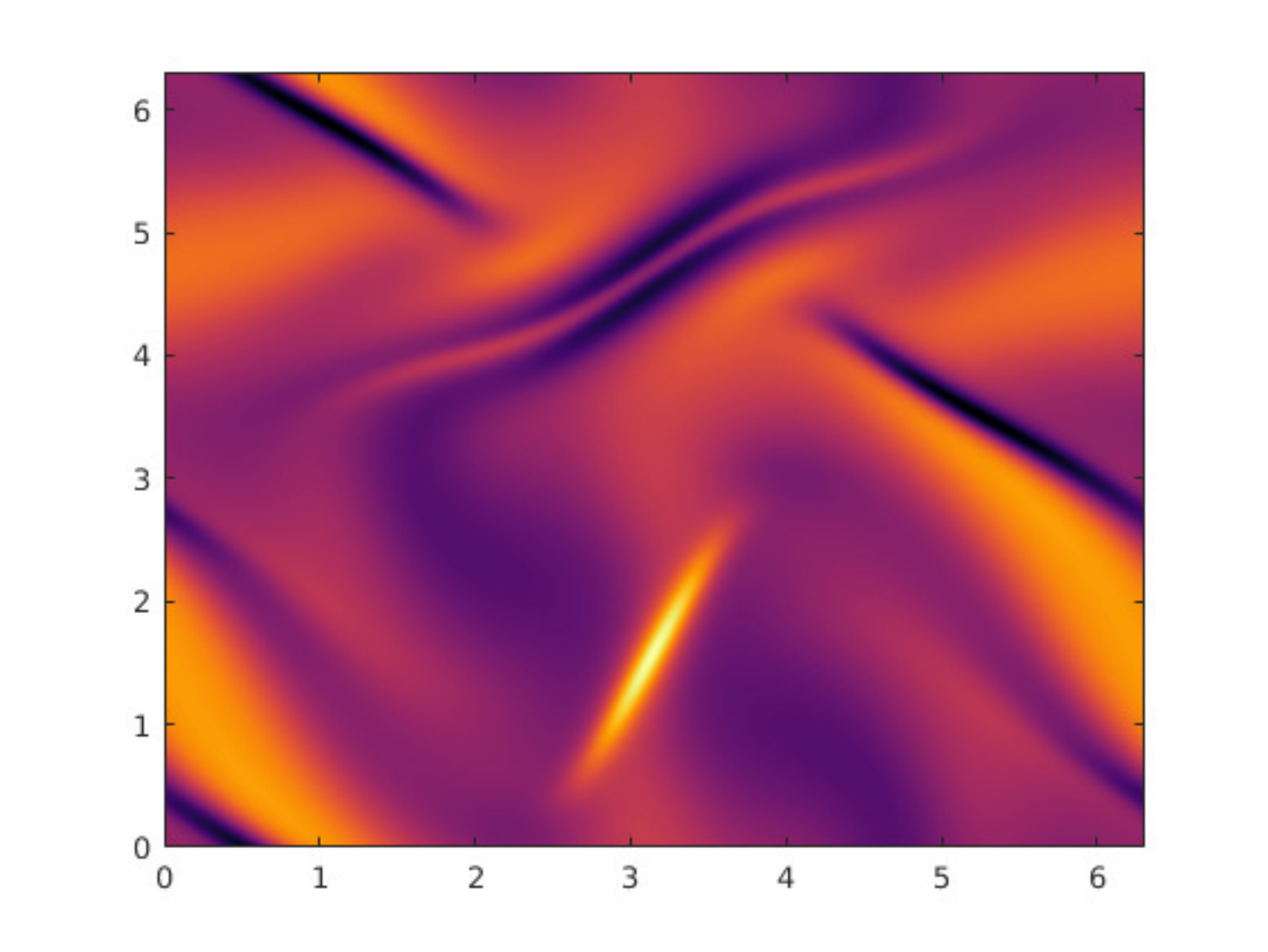} \includegraphics[scale=0.34]{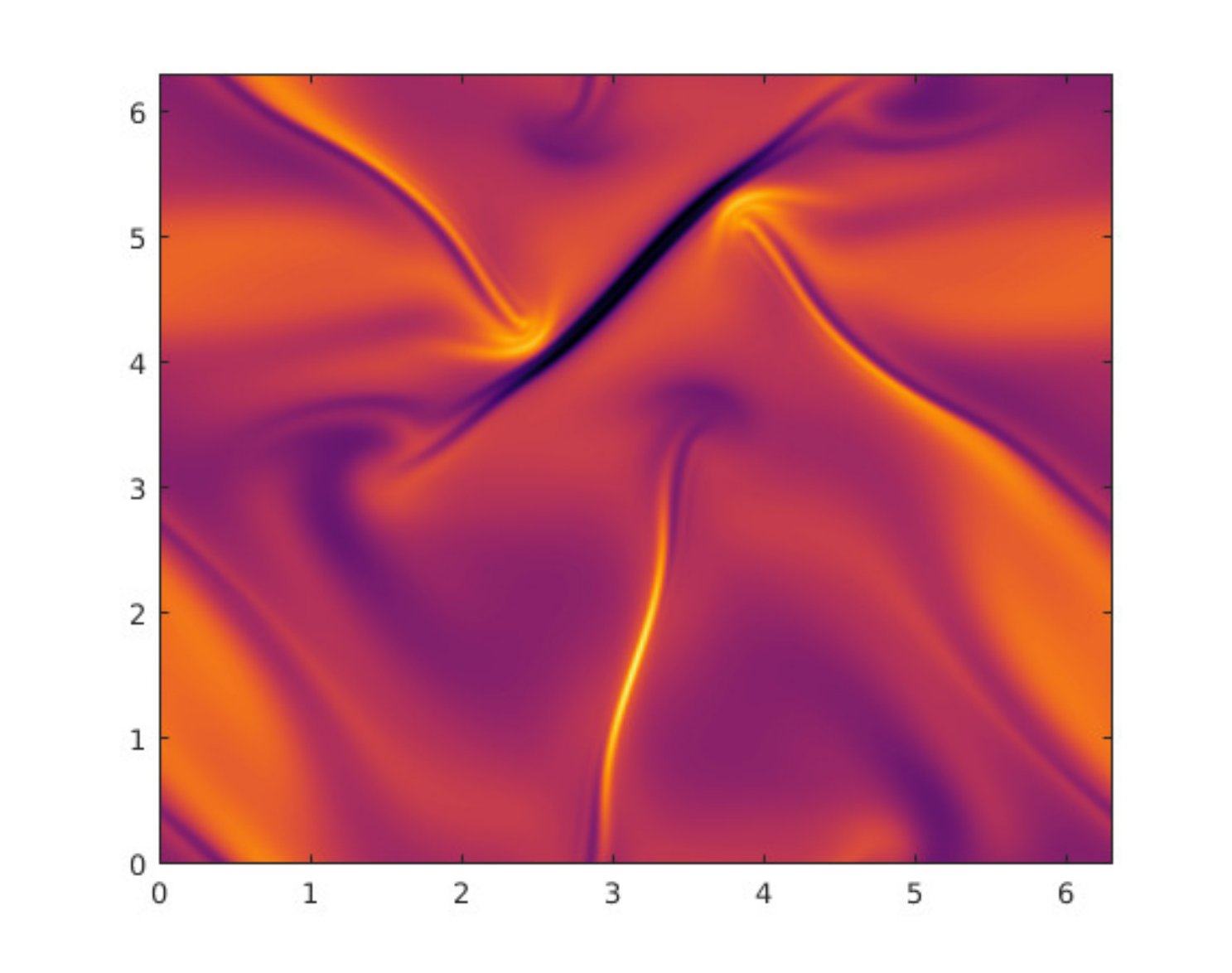}\\
$t=0.4$ \hspace{4cm} $t=0.8$  \hspace{4cm}  $t=1.2$
  \vspace{-2mm}
\caption{Time evolution of $-\Delta\psi$ (current density). \label{fig_2}}
  \vspace{-2mm}
\end{figure}
\begin{figure}
\centering
\includegraphics[scale=0.4]{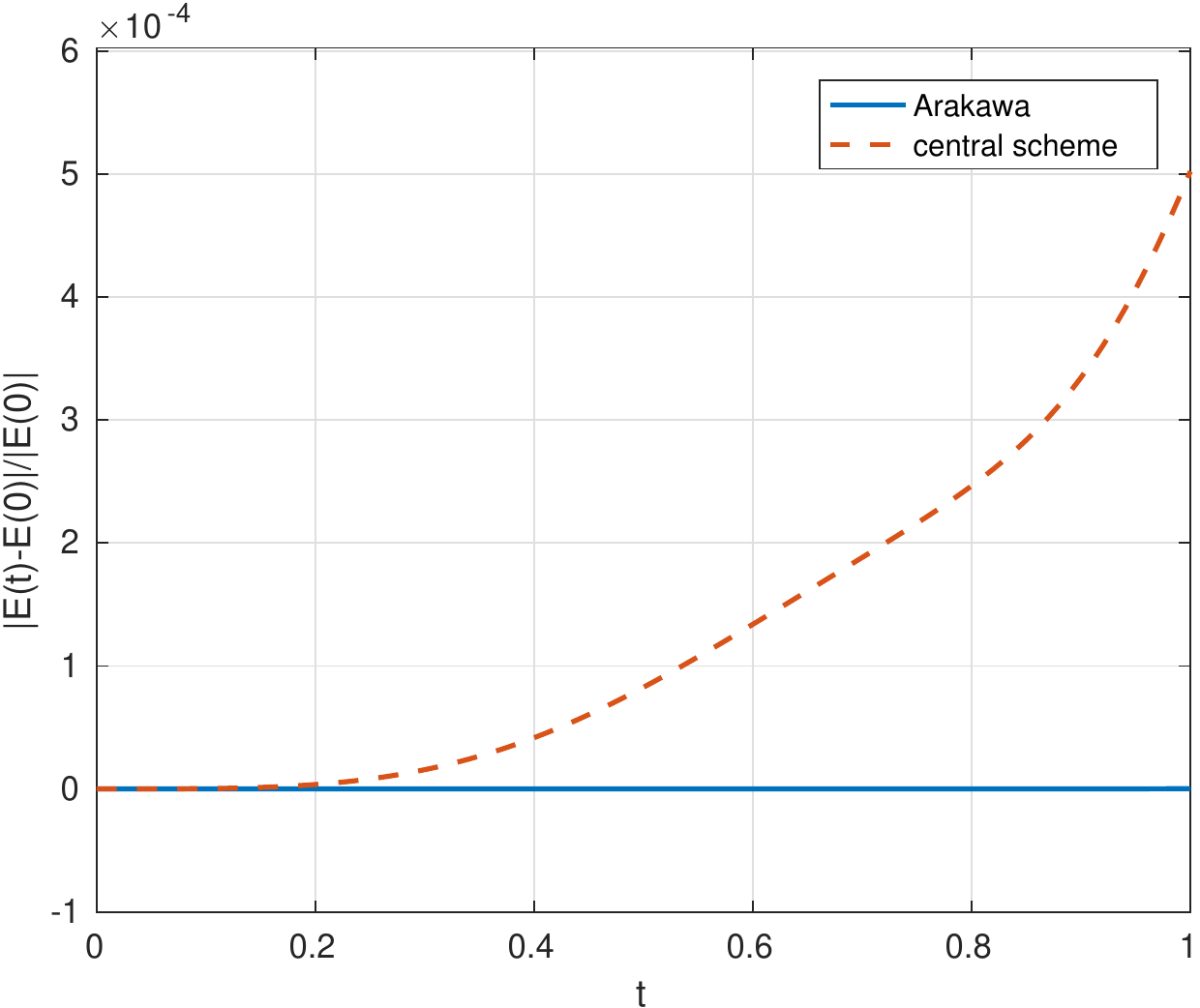} \includegraphics[scale=0.377]{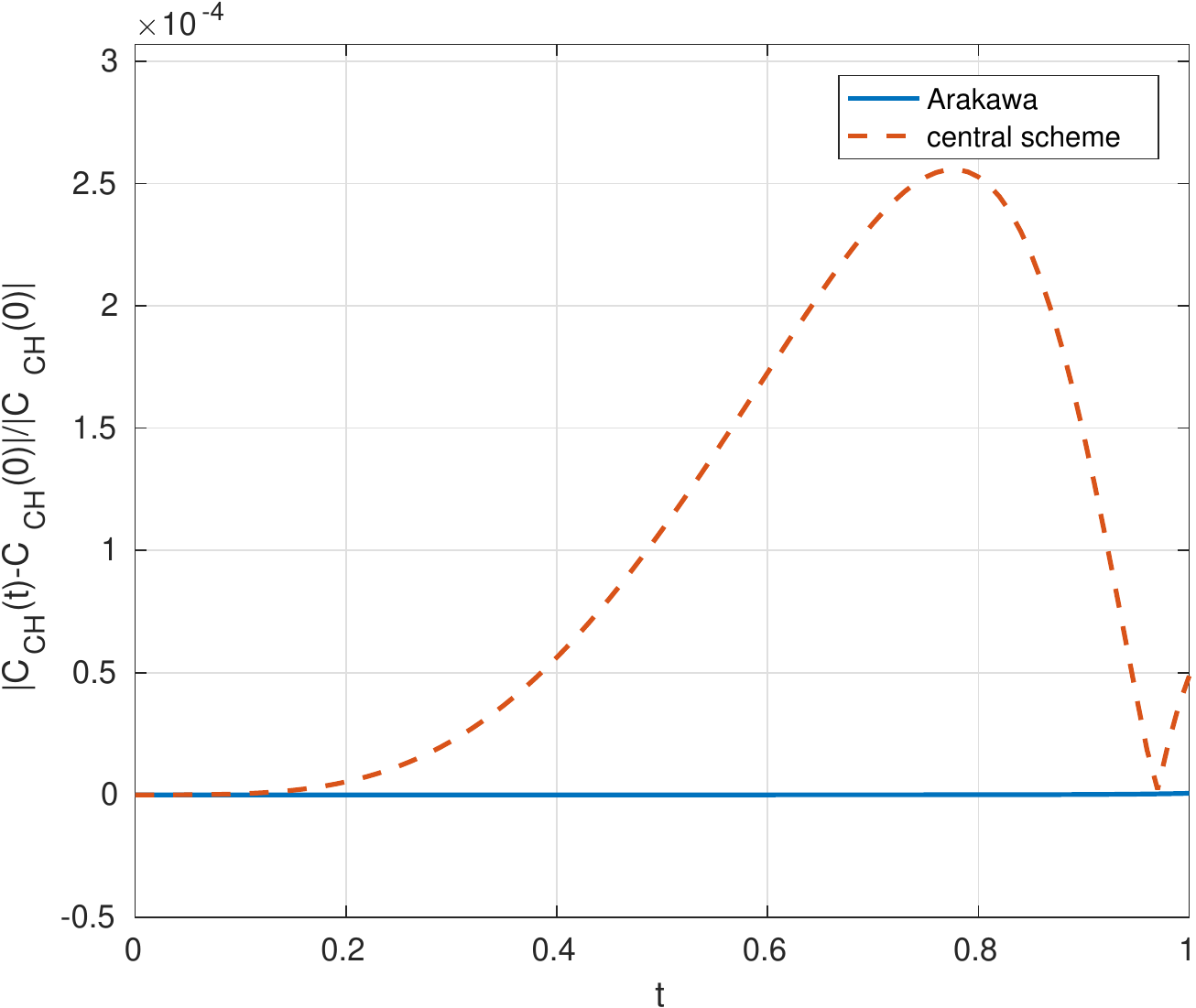} \includegraphics[scale=0.4]{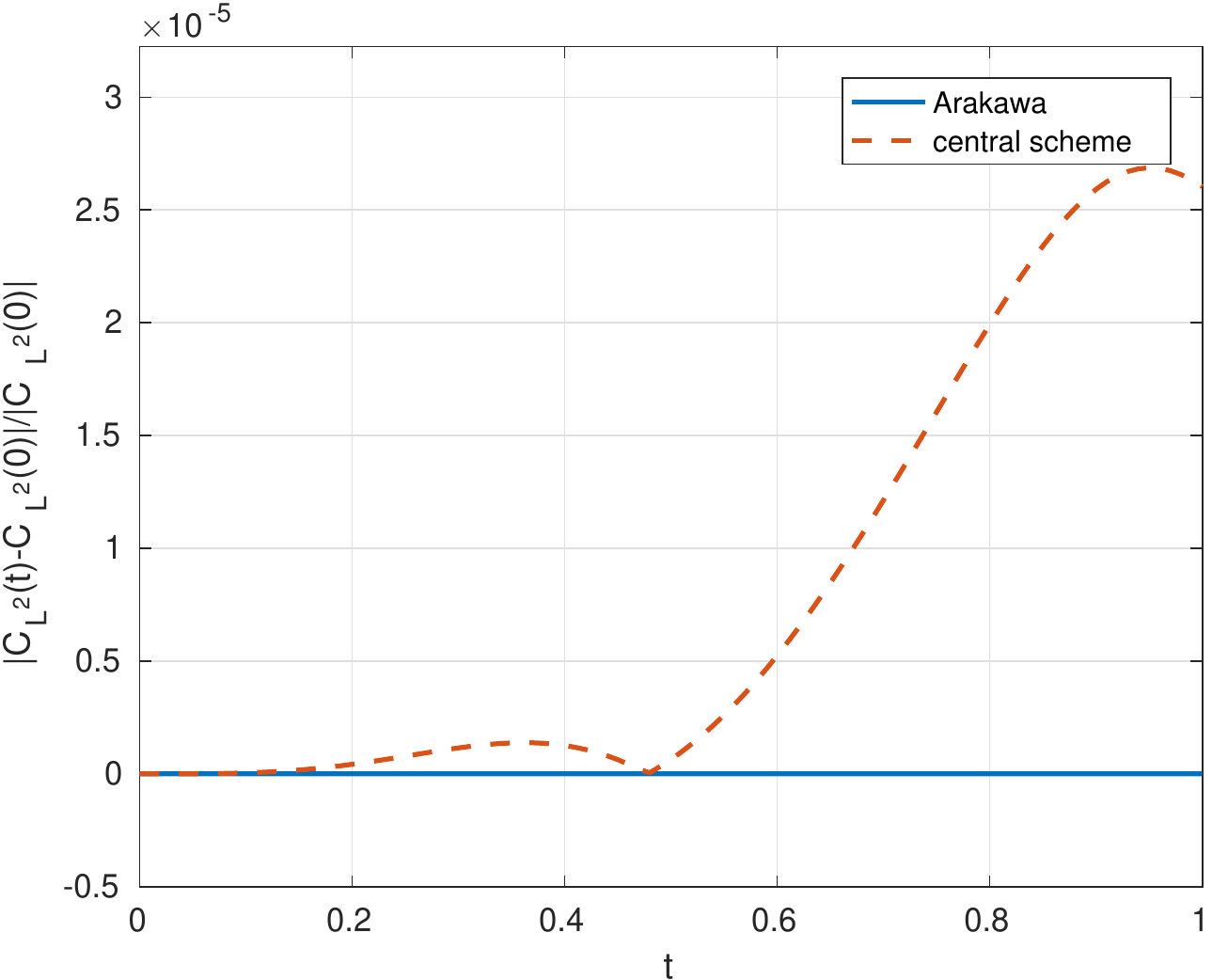}
  \vspace{-2mm}
\caption{The time evolution of the errors of the total energy (left), cross-helicity (center) and the $L^2$ norm of $\psi^*$ (right) for a simple central difference approximation and the Arakawa scheme. \label{fig_3}}
  \vspace{-2mm}
\end{figure}
To illustrate the efficacy of the scheme described above in conserving the energy and the Casimirs we perform a numerical experiment. We restrict ourselves in the simpler case of planar dynamics, i.e. we disregard dynamics parallel to the direction of symmetry simulating the evolution of the Orszag-Tang vortex influenced by the inclusion of electron inertia. The corresponding semi-discrete equations assume the form 
  \begin{eqnarray}
 \frac{d\psi^{*}_{ij}}{dt}= \Jc_{ij}(\chi, \psi^*)\,, \quad 
 \frac{d\omega_{ij}}{dt}=\Jc_{ij}(\chi,\omega)+\Jc_{ij}(\psi,\Delta \psi)\,,\label{discrete_ts_xmhd_dyn_4}
 \end{eqnarray} 
with $\psi^{*}_{ij}=\psi_{ij}-d_e^2(\Delta\psi)_{ij}$. For the time integration we use the explicit RK4 method with time step $h_t=0.01$, while the spatial resolution is $256\times 256$. %For the Poisson and Helmholtz equations, which emerge in the computation of $\chi$ and $\psi$, respectively, we use simple FFT solvers. 
Some snapshots of the vortex evolution with $d_e=0.2$ are depicted in figure \ref{fig_2}. The Arakawa scheme results in errors $|u(t)-u(0)|/|u(0)|$ of the order of $10^{-9}-10^{-8}$ for the discrete analogues of the Hamiltonian $\Hc=0.5\int d^2x \left[\chi \omega -\psi\Delta\psi+d_e^2(\Delta\psi)^2\right]$, the cross-helicity\, $\Cc_{CH}=\int d^2x\, \omega\psi^*$ and the $L^2$ norm of $\psi^*$, $\Cc_{L^2}=\int d^2x\, \psi*^2$. It is clear that this scheme performs far better than the simple central approximation for $\Jc$ (see figure \ref{fig_3}), however, still does not reach the standard of the VI approach adopted in \cite{Kraus2016}, where the implicit Crank-Nicolson method was utilized for the time integration. The error can only come from the time stepping scheme since by construction $\Hc_d$ and $\Zc_d$ are exactly preserved by the spatial discretization. Hence, we expect that the conservation properties can be improved further upon adopting an implicit scheme.  
 \vspace{-2mm}
\section{Summary} 
In this report, a trilinear bracket description of the incompressible 3D XMHD dynamics and also the corresponding description for the translationally symmetric version of the model are presented. Employing a discretization algorithm, which has been introduced in the context of fluid dynamics, we derived a conservative, semi-discrete set of equations that involve the well known Arakawa Jacobian. Subsequently, employing temporal discretization we performed a numerical simulation of planar dynamics and confirmed the good performance of the scheme. This study can serve as a starting point for the construction of a conservative XMHD algorithm with dynamics parallel to the direction of symmetry and can be extended further by employing different methods for spatial discretization, e.g. FEM, that is more efficient for the discretization of realistic domains and also implicit time integration schemes, which exhibit certain merits in terms of stability and conservation properties.
 \vspace{-2mm}
 \section*{Acknowledgements}
 This work has been carried out within the framework of the EUROfusion Consortium and has received funding from the Euratom research and training programme 2014-2018 and 2019-2020 under grant agreement No 633053 as well as from the National Programme for the Controlled Thermonuclear Fusion,
Hellenic Republic. The views and opinions expressed herein do not necessarily reflect those of the European Commission.
The authors would like to thank Dr. Daniela Grasso for useful discussions. DAK and GNT warmly acknowledge the hospitality of the Numerical Plasma Physics Division of the Max Planck Institute for Plasma Physics, Garching, where a part of this research was done.

\end{document}